\documentclass[useAMS,usenatbib]{mn2e}
\usepackage{graphicx}
\usepackage{amsmath,amsfonts,amssymb}
\usepackage{wrapfig}
\usepackage{epsfig}
\usepackage{psfrag}
\usepackage{subfigure}
\usepackage{pstricks,pst-plot} 

\begin{document}

\title[NSBH binaries]
{Periastron Advance in Neutron Star - Black Hole Binaries}

\author[Bagchi]
{\parbox[t]{\textwidth}{Manjari Bagchi$^{1}$\thanks{Email: Manjari.Bagchi@mail.wvu.edu}}\\
\vspace*{3pt} \\
\\ $^1$ Department of Physics, White Hall, West Virginia University, Morgantown, WV 26506, USA
}

\maketitle

\begin{abstract}

As neutron star - black hole binaries are expected to be discovered through future pulsar surveys using upcoming facilities, it is necessary to understand various observable properties of such systems. In the present work, we study the advance of the periastron of such binaries under the post-Newtonian formalism over a wide range of parameters. We find that the first and second order post-Newtonian effects and the leading order spin-orbit coupling effects are significant for such binaries but higher order effects can be neglected. 

\end{abstract}


\section{Introduction} 
\label{lb:introductions} 

Relativistic binary pulsars are very useful tools as they lead to the test of general relativity, accurate determination of the masses of the members of the binary, understanding of the stellar evolution, etc \citep{st03, st04}. So far, such systems include only neutron star - white dwarf and neutron star - neutron star systems. No neutron star - stellar mass black hole (NSBH) system is known yet, although a number of theoretical studies exist \citep[and references therein]{ppr05, kh09, fl11}, discussing the expected number and properties of NSBH binaries formed either through normal evolutionary channel or through exchange interactions in dense stellar environments. NSBH systems are expected to be discovered through future radio pulsar surveys using upcoming facilities like SKA. Moreover, gravitational waves emitted by these systems will be detected soon by the ground based detectors like LIGO, VIRGO, TAMA300, and GEO600. For all these reasons, theoretical efforts to understand the properties of NSBH systems in better details is ongoing. These systems are usually modelled under the Post-Newtonian (PN) formalism. In the present work, we explore which terms in this formalism are more important and deserve more attention. We confine ourselves with only one parameter, namely the advance of the periastron of the orbit of the neutron star, which is a measurable quantity when the neutron star is a radio pulsar. 

The rest of the paper is organized as follows, in Section \ref{lb:anal} we present our study, then in Section \ref{lb:disc} we discuss some other relativistic effects in such systems, and finally our conclusions are in Section \ref{lb:con}.

\section{Analysis} 
\label{lb:anal} 
PN approximation is effectively an expansion of Einstein's theory in powers of a small parameter $\epsilon$ which becomes $(v/c)^2 \sim GM/(c^2 R)$ in case of a compact binary where $v$, $M$, $R$, $G$, and $c$ are the characteristic orbital velocity, the total mass, the typical orbital separation, the gravitational constant, and the speed of light. In addition to conventional PN parameters, spins of the members of the binary also contribute to the dynamics of the system and the total Hamiltonian becomes 
\begin{eqnarray}
H = H_{\rm PN} + H_{\rm SO}^{\rm LO} + H_{S_1 S_2}^{\rm LO} + H_{S_1^2}^{\rm LO} + H_{S_2^2}^{\rm LO} + H_{\rm SO}^{\rm NLO} + H_{S_1 S_2}^{\rm NLO} \nonumber \\
 H_{S_1^2}^{\rm NLO} + H_{S_2^2}^{\rm NLO} +  \ldots {\rm ~higher~ order~ spin ~ terms}. 
\label{eq:k_intro}
\end{eqnarray} where PN stands for the Post-Newtonian expansion, $n$th order term involving $(v/c)^{2n}$. SO is for the spin-orbit coupling, $S_1 S_2$, $S_1^2$ and $S_2^2$ are for the spin-spin interactions. LO means the leading order, NLO stands for the next leading order, NNLO means the next-to-next leading order and so on. The Hamiltonian is known upto 3.5PN and NNLO-SO and NNLO-$S_1 S_2$ \citep[and references therein]{ths10, kfs03,hs11a,hs11b,shs08}, but the observable parameters, e.g., the rate of the periastron advance of the binary are known at most upto 3PN and LO-SO \citep{ds88, kg05}. 

\citet[hereafter KG05]{kg05} parametrized the orbital motion of a compact binary up to 3PN and LO-SO, with a parameter $\tilde{k}$ which becomes the measure of the periastron advance of the binary, i.e. $\tilde{k} = k = \frac{1}{n} \langle  \dot{\omega}  \rangle $, for the non-spinning part. Here $\dot{\omega}$ is the rate of the periastron advance and $n$ is the mean orbital motion which can be expressed in terms of the reduced energy ($E$), the reduced orbital angular momentum ($L$), and $\eta = m_1 \, m_2 /(m_1 + m_2)^2$, $m_1$ and $m_2$ being the masses of the members of the binary (see Eqn. 5.6c of KG05). But $\tilde{k}_{\rm SO}^{\rm LO} \neq k_{\rm SO}^{\rm LO} $ in their formalism. Following KG05, $\tilde{k}$ up to 3PN can be written as:

\begin{equation}
\tilde{k}_{\rm 1PN} = {k}_{\rm 1PN} = \frac{3}{c^2  L^2} ,
\label{eq:k1pn}
\end{equation} 

\begin{equation}
\tilde{k}_{\rm 2PN} = {k}_{\rm 2PN} = \frac{3}{c^2  L^2} \frac{-2E}{4 c^2} \left( -5 + 2 \eta + (35 - 10 \eta)x  \right) ,
\label{eq:k2pn}
\end{equation} and

\begin{eqnarray}
\tilde{k}_{\rm 3PN} = {k}_{\rm 3PN} = \frac{3}{c^2  L^2} \frac{(-2E)^2}{384 c^4} \big[ 120 -120 \eta + 96 \eta^2 + \nonumber \\ (-10080+13952 \eta - 123 \pi^2 \eta -1440 \eta^2) x + \nonumber \\ (36960-40000 \eta + 615 \pi^2 \eta + 1680 \eta^2)x^2 \big],
\label{eq:k3pn}
\end{eqnarray} where $x=1/(-2EL^2)$ and ${k}_{\rm PN} = {k}_{\rm 1PN}+ {k}_{\rm 2PN} +{k}_{\rm 3PN}$ (neglecting higher order PN terms). In Newtonian dynamics, $x=1/(1-e^2)$ where $e$ is the eccentricity of the orbit giving $x \geqslant 1$. The 3PN expressions for $E$ and $L$ are given by \citet{mgs04} in terms of different orbital elements, and here it is possible for $x$ to be smaller than 1, but it can have only such values that the square of the eccentricities are either zero or positive numbers (Eqns. 20b and 20d of \citet{mgs04}). We have checked for a  wide range of values of $(-2E) / (4 c^2)$ that the square of the ``radial eccentricity" ($e_r^2$) remains greater than or equal to zero  for $x \gtrsim 0.59$. So we use only the values of x greater than or equal to $0.59$. 

Eqn. (\ref{eq:k2pn}) implies that ${k}_{\rm 2PN}$ is zero at $x = (5 - 2 \eta) / (35 - 10 \eta) $. For realistic ranges of $\eta$ for NSBH systems, this happens around $x = 0.14$. Here by `realistic ranges of $\eta$' we mean realistic ranges of masses for neutron stars ($m_{ns} = 1-2~M_{\odot}$) and black holes ($m_{bh} = 5-15~M_{\odot}$)\footnote{We will discuss more about these mass ranges later.}. In this range, we choose three representative values of $\eta$, e.g., $\eta=0.107725$ for $m_1 = m_{ns} = 1.4 \, M_{\odot}$ and $m_2 = m_{bh} = 10 \, M_{\odot}$, giving ${k}_{\rm 2PN}=0$ at $x=0.141042 $, for $\eta=0.204082$ ($m_{ns} = 2.0 \, M_{\odot}$, $m_{bh} = 5 \, M_{\odot}$) ${k}_{\rm 2PN}=0$ at $x= 0.139319$, and for $\eta=0.058594$ ($m_{ns} = 1.0 \, M_{\odot}$, $m_{bh} = 15 \, M_{\odot}$) ${k}_{\rm 2PN}=0$ at $x= 0.141884$. Similarly, for specific values of $x$, ${k}_{\rm 3PN}$ becomes zero. As Eqn. (\ref{eq:k3pn}) is quadratic in powers of $x$, for each value of $\eta$, there exist two values of $x$ at which ${k}_{\rm 3PN}$ becomes zero. As examples, for $\eta=0.058594$, ${k}_{\rm 3PN}=0$ at $x=0.254246, 0.012740$, for $\eta=0.107725$, ${k}_{\rm 3PN}=0$ at $x= 0.248754, 0.013051$, and for $\eta=0.204082$, ${k}_{\rm 3PN}=0$ at $x = 0.236489, 0.0139767$. As we have already mentioned that $x$ cannot be less than 0.59, both ${k}_{\rm 2PN}$ and ${k}_{\rm 3PN}$ are always non-vanishing and positive, so they add up in the total ${k}_{\rm PN}$. That is why, it is worthwhile to study the significance of ${k}_{\rm 3PN}$ in comparison with ${k}_{\rm 2PN}$.

\begin{figure}
\centerline{\psfig{figure=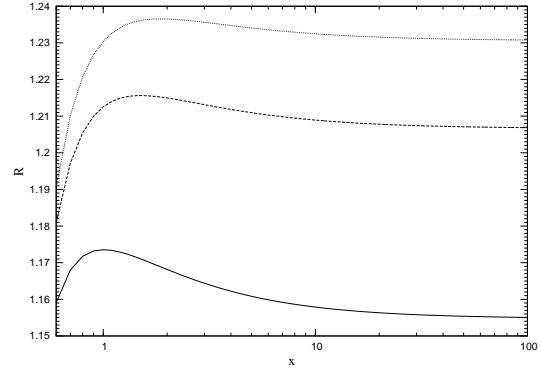,width=5.0cm,angle=270}}
\caption{Variation of $R$ with $x$ for an NSBH binary with $\eta$ values 0.204082 (solid line), 0.107725 (dashed line) and 0.058594 (dotted line) respectively, keeping $x \geq 0.59$.} 
\label{fig:3pn2pncomp}
\end{figure}

Expressions of ${k}_{\rm 1PN}$ and ${k}_{\rm 2PN}$ in terms of basic orbital elements like orbital periods, eccentricities and masses of the components are simple and were first given by \citet[hereafter DS88]{ds88}. On the other hand, expressing ${k}_{\rm 3PN}$ in terms of only those elements is rather difficult. As an alternative simple approach, we decide to compute the ratio 

\begin{eqnarray}
R =  \frac{{k}_{\rm 3PN}/{k}_{\rm 2PN}}{{k}_{\rm 2PN}/{k}_{\rm 1PN}} 
= \frac{1}{24} \frac{1}{ \left[ -5 + 2 \eta + (35 - 10 \eta)x \right] ^2} \times \nonumber \\ 
\left[( 120 -120 \eta + 96 \eta^2 ) + \right.  \nonumber \\  (-10080+13952 \eta - 123 \pi^2 \eta -1440 \eta^2) x   \nonumber \\ 
+ \left. (36960-40000 \eta + 615 \pi^2 \eta + 1680 \eta^2)x^2 \right],
\label{eq:ratio321pn}
\end{eqnarray} which depends only on $\eta$ and $x$. In Fig \ref{fig:3pn2pncomp} we plot $R$ against $x$, $x$ being in the range of 0.59 $-$ 100, for the above mentioned values of $\eta$. We see that $R \sim 1.2$, implying ${k}_{\rm 3PN}/{k}_{\rm 2PN} \simeq   {k}_{\rm 2PN}/{k}_{\rm 1PN}$.

${k}_{\rm 1PN}$ and ${k}_{\rm 2PN}$ are given in terms of orbital elements in DS88 as:
\begin{equation}
{k}_{\rm 1PN} = \frac{3 \beta_0^2 }{1-e_{t}^{2}}, 
\label{eq:k1_terms}
\end{equation}
\begin{equation}
{k}_{\rm 2PN} = \frac{3 \beta_0^4 \, f_0 }{1-e_{t}^{2}} , 
\label{eq:k2_terms}
\end{equation} where 
\begin{equation}
\beta_0~=~\frac{(G M n_0)^{1/3}}{c} ,
\label{eq:beta0}
\end{equation} and
\begin{eqnarray}
f_0~=~\frac{1}{1-e_{t}^{2}}\left( \frac{39}{4}x_{ns}^2+\frac{27}{4}x_{bh}^2+15 x_{ns} x_{bh} \right) \nonumber \\ - \left( \frac{13}{4}x_{ns}^2+\frac{1}{4}x_{bh}^2+\frac{13}{3} x_{ns} x_{bh} \right).
\label{eq:f0}
\end{eqnarray} 
Here $e_{t}$ is the proper time eccentricity \citep{dd85, dd86} which becomes the ordinary eccentricity in case of Newtonian dynamics, $n_0 = 2\pi/P_{orb}$, $P_{orb}$ is the orbital period of the binary, $M = m_{ns} + m_{bh}$, $x_{ns} = m_{ns} / M $, $x_{bh} = m_{bh} / M $, and $\dot{\omega}_{\rm 1,2PN} = n_0 \, \left( k_{\rm 1PN} + k_{\rm 2PN} \right)$.

\begin{figure}
\centerline{\psfig{figure=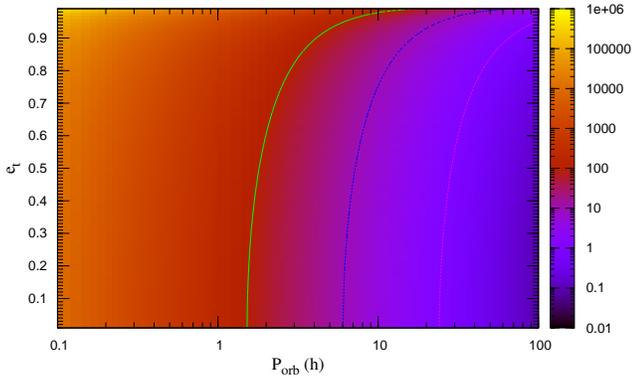,width=9.2cm,angle=0}}
\caption{Variation of $\dot{\omega}_{\rm 1PN}$ with orbital periods and orbital eccentricities for an NSBH binary having $m_{ns} = 1.4 \, M_{\odot}$, $m_{bh} = 10 \, M_{\odot}$. X and Y axes represent orbital periods in hours and orbital eccentricities respectively while the color code represents the values of $\dot{\omega}_{\rm 1PN}$ in ${\rm deg~ yr^{-1}}$. The contours of $\dot{\omega}_{\rm 1PN} = 100$, $10$, and $1$ ${\rm deg~ yr^{-1}}$ are also shown from left to right.} 
\label{fig:omdot1pn}
\end{figure}

For double neutron star (DNS) systems, observed $\dot{\omega}$ is modelled using the 1PN term only, which is justified, as even for the most relativistic system PSR J0737-3039, $k_{\rm 2PN}/k_{\rm 1PN} = 2.6 \times 10^{-5}$, but the 2PN term should be considered if a more relativistic DNS system is discovered. The observed value of $\dot{\omega}$ ($16.899~{\rm deg ~ yr^{-1}}$) for PSR J0737-3039 \citep{ks06} is the maximum among all known DNSs. For the sake of comparison, in Fig. \ref{fig:omdot1pn}, we show the variation of $\dot{\omega}_{\rm 1PN} = n_0 \, k_{\rm 1PN}$ with orbital periods and eccentricities for an NSBH binary having $m_{ns} = 1.4 \, M_{\odot}$, $m_{bh} = 10 \, M_{\odot}$. The X and Y axes represent orbital periods in hours and orbital eccentricities respectively while the color code represents the values of $\dot{\omega}_{\rm 1PN}$ in ${\rm deg~ yr^{-1}}$. The contours of $\dot{\omega}_{\rm 1PN} = 100$, $10$, and $1$ ${\rm deg~ yr^{-1}}$ are also shown from left to right. Here one should remember that for a relativistic binary, both $P_{orb}$ and $e_{t}$ decrease with time due to the emission of gravitational waves. As an example, for an NSBH binary having $m_{ns} = 1.4 \, M_{\odot}$, $m_{bh} = 10 \, M_{\odot}$, $P_{orb} = 10$ hours and $e_t = 0.3$, 1PN values are $\dot{P}_{orb} = - 1.04 \times 10^{-12}~{\rm s~s^{-1}}$ and $\dot{e}_{t} = - 6.73 \times 10^{-18} ~{\rm s^{-1}}$ \citep{pet64}.

\begin{figure}
\centerline{\psfig{figure=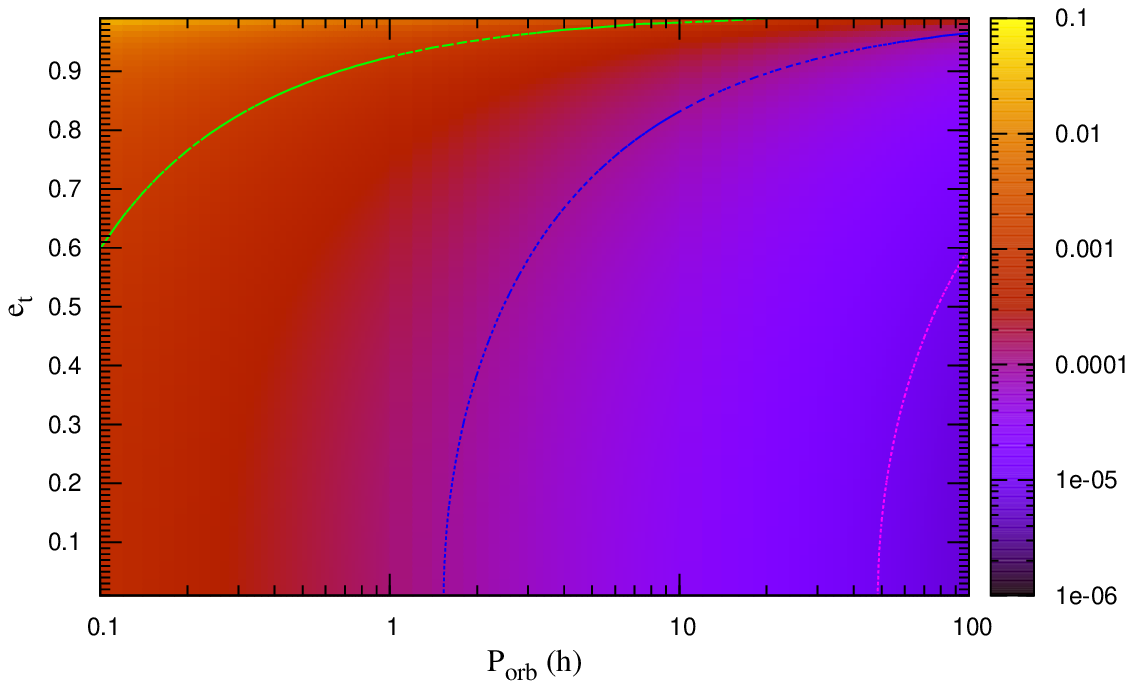,width=9.2cm,angle=0}}
\caption{Variation of ${k}_{\rm 2PN}/{k}_{\rm 1PN}$ with orbital periods and orbital eccentricities for an NSBH binary having $m_{ns} = 1.4 \, M_{\odot}$, $m_{bh} = 10 \, M_{\odot}$. X and Y axes represent orbital periods in hours and orbital eccentricities respectively while the color code represents the values of ${k}_{\rm 2PN}/{k}_{\rm 1PN}$. The contours of ${k}_{\rm 2PN}  / {k}_{\rm 1PN} = 10^{-3}$, $10^{-4}$, and $10^{-5}$ are also shown from left to right.} 
\label{fig:2pn1pncomp}
\end{figure}

In Fig \ref{fig:2pn1pncomp} we plot ${k}_{\rm 2PN}/{k}_{\rm 1PN}$ for the same NSBH binary. The X and Y axes have the same meaning and the color code represents the values of ${k}_{\rm 2PN}/{k}_{\rm 1PN}$. The contours of ${k}_{\rm 2PN}/{k}_{\rm 1PN} = 10^{-3}$, $10^{-4}$ and $10^{-5}$ are also shown from left to right. As the phase-space with very high eccentricities and very short orbital periods is unlikely\footnote{Such systems are unlikely to form and even if they form under some extra-ordinary situation, they would merge very soon making the probability of detection very small.}, we can say ${k}_{\rm 2PN}/{k}_{\rm 1PN} < 10^{-3}$. As we have already seen that ${k}_{\rm 3PN}/{k}_{\rm 2PN} \sim {k}_{\rm 2PN}/{k}_{\rm 1PN}$, it can be concluded that ${k}_{\rm 3PN}/{k}_{\rm 1PN} \lesssim 10^{-6}$. Since the accuracy as good as $10^{-5}$ in the measurement of $\dot{\omega}$ has been already achieved for many binary pulsars, one might aim for the same for NSBH binaries. In that case, one must consider 2PN effects, but 3PN and higher order PN effects can be excluded unless further accuracy is intended.

Now, we wish to study the effect of the leading order spin-orbit coupling. According to DS88,

\begin{equation}
{k}_{\rm SO}^{\rm LO} = - \frac{3 \beta_0^3}{1-e_{t}^2}\left(g_{s,ns} \beta_{s,ns}+g_{s,bh} \beta_{s,bh} \right) ,
\label{eq:k_so}
\end{equation} where

\begin{equation}
\beta_{s,ns}~=~\frac{c I_{ns} }{G m_{ns}^2} \cdot \frac{2 \pi}{P_{s,ns}}
\label{eq:betasns}
\end{equation} and

\begin{equation}
\beta_{s,bh}~=~\frac{c^2 a_{bh} }{G m_{bh}} \, .
\label{eq:betasbh}
\end{equation} 
Here $I_{ns}$ is the moment of inertia of the neutron star, $a_{bh}= ( 2 \pi I_{bh} ) / ( c \, m_{bh} \, P_{s, bh} )$ is the spin parameter of the black hole ($I_{bh}$ is the moment of inertia of the black hole and $P_{s, bh}$ is its spin period), $P_{s,ns}$ is the spin period of the neutron star. It is well known that $a_{bh} \leqslant G m_{bh} / c^2$, with $a_{bh, max} = G m_{bh} / c^2$ giving the ``maximally rotating black hole". From Eqns (\ref{eq:k1_terms}) and (\ref{eq:k_so}), we can write

\begin{equation}
W = \bigg| \frac{{k}_{\rm SO}^{\rm LO} } {{k}_{\rm 1PN}} \bigg| =  \left(g_{s,ns} \beta_{s,ns}+g_{s,bh} \beta_{s,bh} \right) \beta_{0} .
\label{eq:k_so1pn}
\end{equation} 

Let us denote ${\bf s_{ns}}$ as the unit spin vector of the neutron star, ${\bf s_{bh} }$ as the unit spin vector of the  black hole, ${\bf k}$ as the unit orbital angular momentum vector, ${\bf h_{ns}}$ as the unit vector from the earth towards the neutron star perpendicular to the plane of the sky, ${\bf h_{bh}}$ as the unit vector from the earth towards the black hole perpendicular to the plane of the sky. Let us also assume that $i$ is the inclination angle of the orbital plane with respect to the sky plane. Then, following DS88,

\begin{eqnarray}
g_{s,ns}~=\frac{x_{ns} \left(4 x_{ns}+ 3x_{bh}\right)}{6(1-e_{t}^2)^{1/2} {\rm sin^2} i}   \left[ (3 ~{\rm sin^2} i-1)~{ \bf   k}+{\rm cos} \, i~ { \bf h_{ns}} \right] {\bf .~ s_{ns} }
\label{eq:gsns}
\end{eqnarray} and

\begin{eqnarray}
g_{s,bh}~=\frac{x_{bh} \left(4 x_{bh}+ 3x_{ns}\right)}{6(1-e_{t}^2)^{1/2} {\rm sin^2} i}  \left[ (3 ~{\rm sin^2} i-1)~{ \bf   k}+{\rm cos} \, i~ { \bf h_{bh}} \right] {\bf .~ s_{bh} }
\label{eq:gsbh}
\end{eqnarray}  

$g_{s,ns}$ will have the maximum value when ${\bf s_{ns}}$ is parallel to the vector $(3~ {\rm sin^2 } i-1)~{ \bf   k}+{\rm cos}~ i~ { \bf h_{ns}} $, giving
\begin{equation}
 {g_{s, ns, ~max}}=~\left[3+\frac{1}{{\rm sin ^2} ~i} \right]^{1/2}\frac{x_{ns} \left(4 x_{ns}+ 3x_{bh}\right)}{6(1-e_{t}^2)^{1/2}} .
\label{eq:gsnsmax}
\end{equation}
For any other orientation of the vectors, $g_{s,ns}$ would be different, but it would always be in the range of $+ {g_{s,ns, ~max}}$ to $- {g_{s,ns, ~max}} $. In case of ${\bf s_{ns}}~\Vert~ {\bf k}$, one obtains
\begin{equation}
g_{s,ns,~\Vert}=~\frac{x_{ns} \left(4 x_{ns}+ 3x_{bh}\right)}{3(1-e_{t}^2)^{1/2}} .
\label{eq:gsns_par}
\end{equation}

Similarly, $g_{s,bh}$ will have the maximum value when ${\bf s_{bh}}$ is parallel to the vector $(3~ {\rm sin^2 } i-1)~{ \bf   k}+{\rm cos}~ i~ { \bf h_{bh}} $, giving
\begin{equation}
 {g_{s,bh, ~max}}=~\left[3+\frac{1}{{\rm sin ^2} ~i} \right]^{1/2}\frac{x_{bh} \left(4 x_{bh}+ 3x_{ns}\right)}{6(1-e_{t}^2)^{1/2}} .
\label{eq:gsbhmax}
\end{equation}
For any other orientation of the vectors, $g_{s,bh}$ would be different, but it would always be in the range of $+ {g_{s,bh, ~max}}$ to $- {g_{s,bh, ~max}}$. In case of ${\bf s_{bh}}~\Vert~ {\bf k}$, one obtains
\begin{equation}
g_{s,bh,~\Vert}=~\frac{x_{bh} \left(4 x_{bh} + 3x_{ns}\right)}{3(1-e_{t}^2)^{1/2}} .
\label{eq:gsbh_par}
\end{equation}

From the above equations, it is clear that ${g_{s,ns, ~max}}$ differs from $g_{s,ns,~\Vert}$ (or ${g_{s,bh, ~max}}$ from $g_{s,bh,~\Vert}$) only by the factor $\frac{1}{2} \left[3+\frac{1}{{\rm sin ^2} ~i} \right]^{1/2}$, which lies between $1.7 - 1.0$ for $i$ in the range of $20^{\circ} - 90^{\circ}$.

To study the spin-orbit coupling effects for an NSBH binary, one needs to know the moment of inertia of the neutron star and the spin parameter of the black hole in addition to their masses, orbital parameters and the spin period of the neutron star. The value of the moment of inertia of a neutron star of known mass depends upon the Equation of State of the matter, which is not very well constrained at present. We choose the values of the moment of inertia as $0.8 \times 10^{38}$, $1.0 \times 10^{38}$, and $1.4 \times 10^{38}~ {\rm kg ~m^2}$ respectively for neutron stars of masses 1.0, 1.4 and 2.0 $M_{\odot}$ according to the predictions of most of the standard Equations of State \citep{ab77, bh02, ls05, mnj10}.

\begin{figure}
\centerline{\psfig{figure=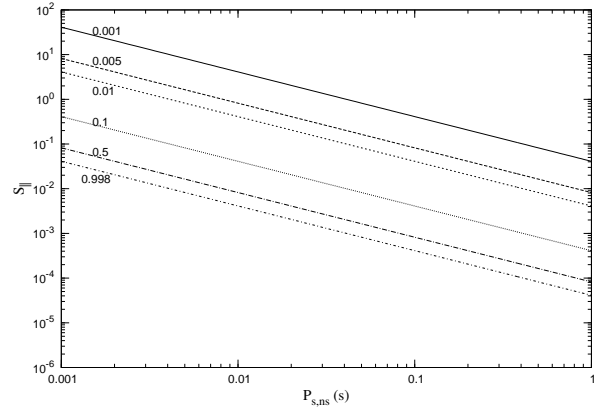,width=8cm,angle=0}}
\caption{Variation of $S_{\Vert}$  with $P_{s, ns}$ for an NSBH binary having $m_{ns} = 1.4 \, M_{\odot}$, $m_{bh} = 10 \, M_{\odot}$. The lines are marked with the corresponding values of $a_{bh}/a_{bh, max}$.} 
\label{fig:nstobhSOcompare}
\end{figure}

First, we compare the spin-orbit coupling due to the neutron star with that due to the black hole by computing the ratio 
\begin{equation}
S_{\Vert} = \frac{ g_{s,ns,~\Vert} \beta_{s,ns} }{ g_{s,bh,~\Vert} \beta_{s,bh} } ~.
\end{equation}
\label{eq:ratioS} 
This ratio is independent of $P_{orb}$ and $e_{t}$. In Fig. \ref{fig:nstobhSOcompare}, we plot $S_{\Vert}$ against $P_{s, ns}$ for the same NSBH binary for different values of $a_{bh}/a_{bh, max}$. Here we use the limit $a_{bh} = 0.998 \, a_{bh, max}$ as given by \citet{thorn74}. It is clear that usually the effect due to the spin of the black hole dominates, i.e. $S_{\Vert} < 1$, unless either the neutron star is rotating very fast or the black hole is rotating very slowly. As some examples, to get $S_{\Vert} > 1$, $a_{bh}$ must be less than $0.01 \, a_{bh, max}$ for $P_{s, ns} = 4$ ms, $a_{bh}$ must be less than $0.005 \, a_{bh, max}$ for $P_{s, ns} = 10$ ms, and $a_{bh}$ must be less than $0.001 \, a_{bh, max}$ for $P_{s, ns} = 40$ ms. 

\begin{figure*}
 \begin{center}
\subfigure[$P_{s, ns} = 0.1$ s, $a_{bh} = 0.001 \, a_{bh, max}$]{\label{subfig:subfig2}\includegraphics[width=0.46\textwidth,angle=0]{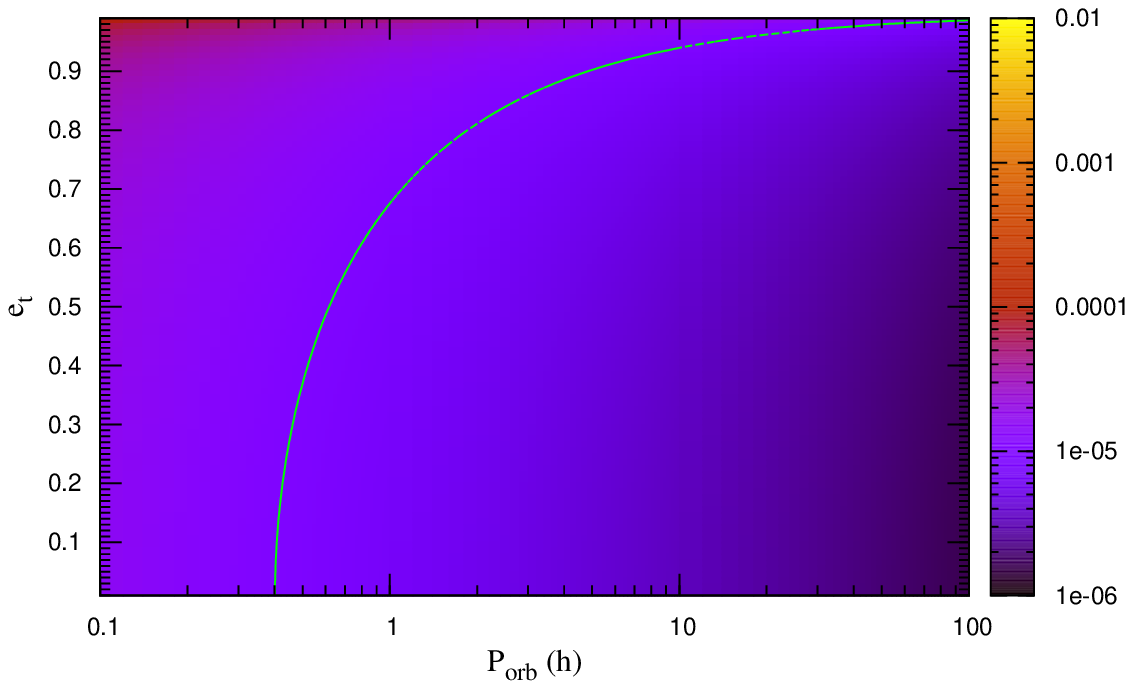}}
\subfigure[$P_{s, ns} = 0.1$ s, $a_{bh} = 0.01 \, a_{bh, max}$]{\label{subfig:subfig1}\includegraphics[width=0.46\textwidth,angle=0]{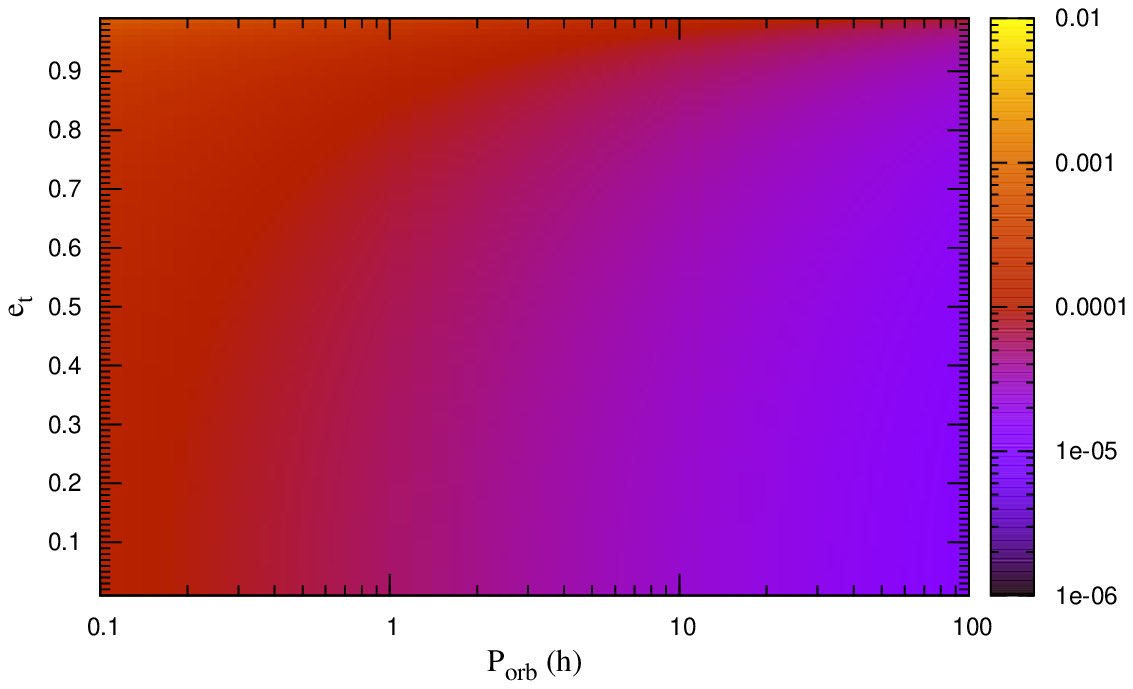}}
\subfigure[$P_{s, ns} = 0.01$ s, $a_{bh} = 0.001 \, a_{bh, max}$]{\label{subfig:subfig4}\includegraphics[width=0.46\textwidth,angle=0]{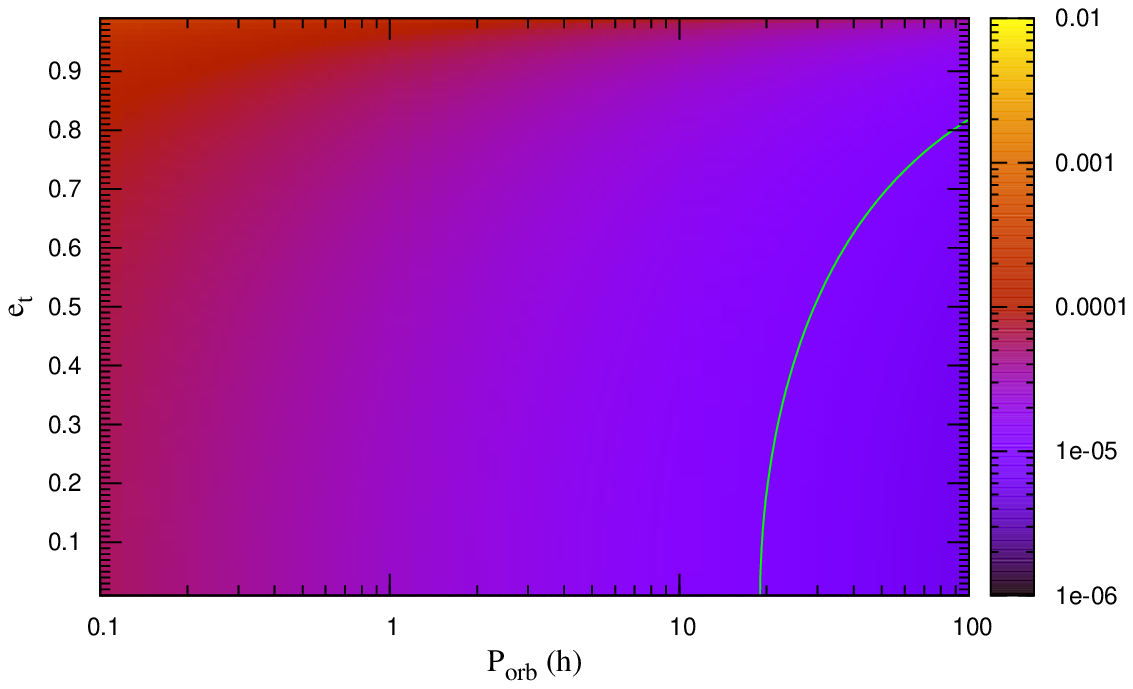}}
\subfigure[$P_{s, ns} = 0.01$ s, $a_{bh} = 0.01 \, a_{bh, max}$]{\label{subfig:subfig3}\includegraphics[width=0.46\textwidth,angle=0]{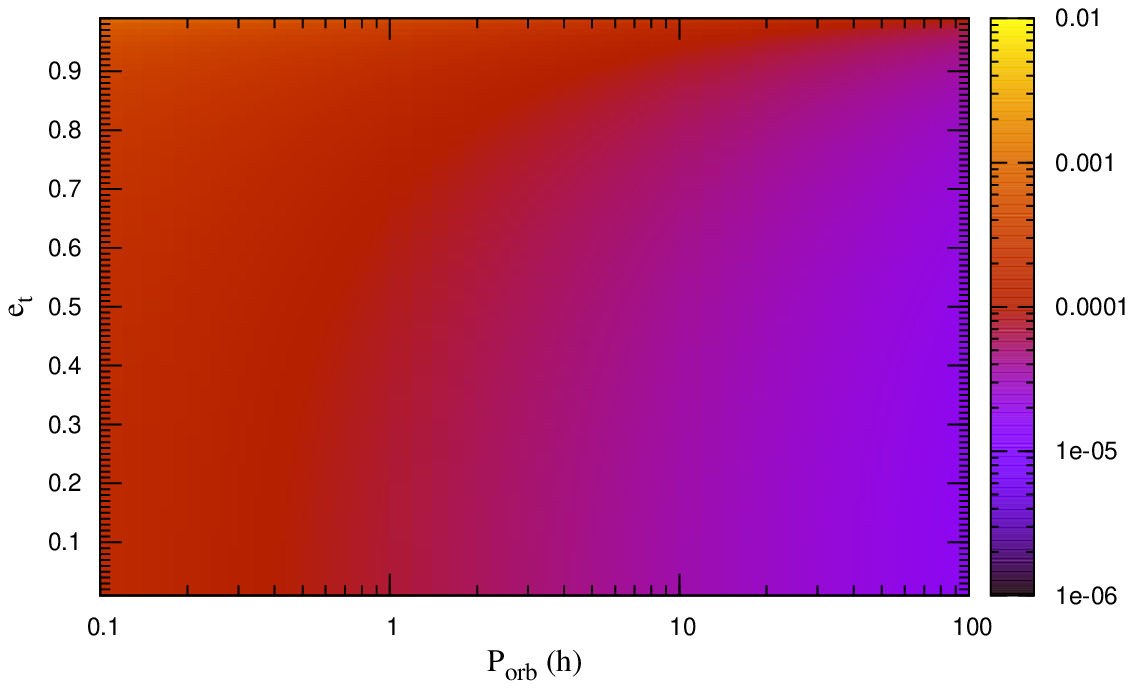}}
 \end{center}
\caption{Variation of $W_{\Vert}$ with orbital periods and orbital eccentricities for an NSBH binary having $m_{ns} = 1.4 \, M_{\odot}$, $m_{bh} = 10 \, M_{\odot}$. X and Y axes represent orbital periods in hours and orbital eccentricities respectively while the color code represents the values of $W_{\Vert}$. The contours for $W_{\Vert} = 10^{-5}$ also have been plotted, whenever in the range.}
\label{fig:soto1pncomp}
\end{figure*}

Now, to answer the question whether the total spin-orbit coupling effect is comparable to 1PN effect, we compute the value of $W_{\Vert}$ where we replace $g_{s,ns}$ by $g_{s,ns,~\Vert}$ and $g_{s,bh}$ by $g_{s,bh,~\Vert}$ in Eqn. (\ref{eq:k_so1pn}).
We plot $W_{\Vert}$ against orbital periods and orbital eccentricities for the same NSBH binary in Fig. \ref{fig:soto1pncomp}. 
As expected, $W_{\Vert}$ increases as either the black hole or the neutron star or both become faster. $W_{\Vert}$ increases with the increase of $e_{t}$ and decreases with the increase of $P_{orb}$ provided all other parameters are unchanged. It is clear that usually $W_{\Vert} > 10^{-5}$ unless the black hole is rotating very slowly, e.g. for $a_{bh} = 0.01 \, a_{bh, max}$, the whole $P_{orb} - e_{t}$ space has $W_{\Vert} > 10^{-5}$ for both $P_{s, ns} = 0.1$ s and $P_{s, ns} = 0.01$ s (even for $P_{s, ns} = 0.001$ s which has not been shown in the figure). For $a_{bh} = 0.001 \, a_{bh, max}$, a significant portion of $P_{orb} - e_{t}$ space has $W_{\Vert} < 10^{-5}$ when $P_{s, ns} = 0.1$ s, but for $P_{s, ns} = 0.01$ s only the region with high $P_{orb}$ and small $e_{t}$ has $W_{\Vert} < 10^{-5}$. These facts imply that $k_{\rm SO}^{\rm LO}$ term might have significant contribution to the observed $\dot{\omega}$ for an NSBH binary depending on the spin and orbital parameters of the system. But here we have assumed that the spin vectors of both the black hole and the neutron star are aligned with the orbital angular momentum vector. In reality, they can be misaligned, specially if the NSBH system was formed through an exchange interaction. For some specific orientations of the spin vectors, combined with appropriate values of other relevant parameters, $k_{\rm SO}^{\rm LO}$ can be very small and even zero. But as only a small window in the total phase-space can lead to this situation, we consider this situation to be unlikely, although not impossible.

In general, $W_{\Vert}$ is not so high that we should expect the non-leading order of the spin-orbit coupling to be effective. As an extreme example, let us consider the NSBH system to have $P_{orb}= 0.5$ hours, $e_{t} = 0.5$, $ a_{bh} = 0.9 \, a_{bh, max}$, and $P_{s, ns} = 1$ ms, giving $W_{\Vert} = 7.16 \times 10^{-3}$. If we assume that LO-SO/1PN $\sim$ NLO-SO/LO-SO, then NLO-SO/1PN becomes $\sim 5 \times 10^{-5}$. In this case, the non-leading order would contribute to the observed value of $\dot{\omega}$. But this set of parameters is not very likely, most of the parameter space would probably lead to NLO-SO/1PN $< 10^{-6}$. Further exploration is needed before making any strong statement, which is beyond the scope of this paper due to the lack of the analytical expression for $k_{\rm SO}^{\rm NLO}$.

So far, we have displayed the results only for a sample NSBH binary with the neutron star of mass $1.4~ M_{\odot}$ and the black hole of mass $10~M_{\odot}$ except for the variation of $R$ with $x$. This choice of the mass of the black hole may seem to be rather high as \citet{opnm10} found that the masses of the known black holes in low mass X-ray binaries have a narrow distribution at $\sim 7.8 ~ M_{\odot}$, population synthesis for NSBH binaries by \citet{ppr05} also predicted the mean mass of the black holes to be around $7 ~ M_{\odot}$. On the other hand, \citet{fsc11} found that the inclusion of high mass systems resulted the mean of the masses of known black holes to be 10-11 $M_{\odot}$. \citet{fl11} used the black hole mass to be $10 ~ M_{\odot}$ in their study of the exchange scenario. Moreover, the higher value of the mass of the black hole makes the system more relativistic. Nevertheless, we have checked that our main conclusions remain unchanged for realistic ranges of the masses, i.e. the neutron star mass in the range of 1-2 $M_{\odot}$ and the black hole mass in the range of 5-15 $M_{\odot}$. We have already seen that, $R$ is always around 1.2 for the above ranges of masses. The effects of the variation of the mass of the neutron star on $k_{2PN}/k_{1PN}$, $S_{\Vert}$ and $W_{\Vert}$ are insignificant, but the variation of the mass of the black hole has small but significant effects. As an example, for $m_{ns} = 1.4~M_{\odot}$, $P_{orb}=10$ hours, $e_{t}=0.3$, the values of $k_{2PN}/k_{1PN}$ are $2.1 \times 10^{-5}$, $3.2 \times 10^{-5}$, and $4.1 \times 10^{-5}$ for $m_{bh}= 5, ~10, ~15 ~M_{\odot}$ respectively. The fact that the phase-space for $k_{2PN}/k_{1PN} > 10^{-3}$ is unphysical, is always true, and thus our main conclusion that  $k_{3PN}/k_{1PN} \lesssim 10^{-6}$ remains unaltered. Similarly, the maximum value of $S_{\Vert}$ in Fig. \ref{fig:nstobhSOcompare} is 41 where $P_{s, ns} = 0.001$ s and $a_{bh}/a_{bh, max} = 0.001$ for $m_{bh} = 10 ~M_{\odot}$, which becomes 87 if $m_{bh} = 5 ~M_{\odot}$ and 27 if $m_{bh} = 15 ~M_{\odot}$ keeping $P_{s, ns}$ and $a_{bh}/a_{bh, max}$ fixed. The minimum value of $S_{\Vert}$ in Fig. \ref{fig:nstobhSOcompare} is $4.1 \times 10^{-5}$ where $P_{s, ns} = 1 $ s and $a_{bh}/a_{bh, max} = 0.998$ for $m_{bh} = 10 ~M_{\odot}$, which becomes $8.7 \times 10^{-5}$ if $m_{bh} = 5 ~M_{\odot}$ and $2.7 \times 10^{-5}$ if $m_{bh} = 15 ~M_{\odot}$ keeping $P_{s, ns}$ and $a_{bh}/a_{bh, max}$ fixed. For $m_{ns} = 1.4~M_{\odot}$, $P_{orb}=10$ hours, $e_{t}=0.3$, $P_{s, ns} = 10$ ms, $a_{bh} = 0.3 \, a_{bh,max}$ the values of $W_{\Vert}$ are $5.6 \times 10^{-4}$, $7.7 \times 10^{-4}$, and $9.2 \times 10^{-4}$ for $m_{bh}= 5, ~10, ~15 ~M_{\odot}$ respectively.

We have considered only the short period NSBH binaries, although very wide ($P_{orb} \gtrsim 1000$ days) NSBH systems can also form \citep{kh09}. We have excluded such wide systems from our study as it is clear that higher order effects would be negligible for such systems. For compact NSBH systems, our chosen ranges for $P_{orb}$ and $e_{t}$ are sufficiently wide to incorporate different formation scenarios, e.g. normal evolutionary formation and the formation through exchange interactions. The same logic applies for the choice of wide ranges for $P_{s, ns}$ and $a_{bh}$.

\section{Discussions} 
\label{lb:disc} 

In this study we have not discussed the effect of the spin-spin interaction. According to \citet{wex95}, spin-spin interaction is significantly smaller than 2PN effect, and can be neglected.

Five post-Keplerian parameters can be measured for a relativistic binary, the rate of the advance of the periastron ($\dot{\omega}$), the rate of change of the orbital period ($\dot{P}_{orb}$), Einstein delay parameter ($\gamma$), Shapiro range parameter ($r$), and Shapiro shape parameter ($s = \sin i$) \citep{dt92,lk05}. We have discussed $\dot{\omega}$ in details, and briefly mentioned about 1PN $\dot{P}_{orb}$. Other effects might also contribute significantly to the observed value of $\dot{P}_{orb}$, like the tail effect and the spin-orbit coupling as discussed by \citet{rs97}. Another interesting hypothesis is the outspiral of the NSBH binary due to the evaporation of the black hole caused by the extra spatial dimensions. According to \citet{skk11}, this effect will be measurable for an NSBH binary and will test the validity of extra dimensions. Lowest order values of $\gamma$ and $r$ for an NSBH system having $m_{ns} = 1.4 \, M_{\odot}$, $m_{bh} = 10 \, M_{\odot}$, $P_{orb} = 10$ hours and $e_t = 0.3$ are $0.013$ s and $4.9 \times 10^{-5}$ s. These are orders of magnitude higher than those for J0737-3039 \citep{ks06}. So it might be worth exploring higher order effects for these parameters in case of NSBH binaries.

In addition to contributing to the periastron advance, spin-orbit coupling also leads to the geodetic precession where both spin vectors precess around the total angular momentum of the system causing changes in the pulse profile. This effect has been observed for some relativistic systems \citep{st03}. The predicted rate of precession upto the leading order \citep{lk05} for an NSBH binary having $m_{ns} = 1.4~{M_{\odot}}$, $m_{bh} = 10~{M_{\odot}}$, $P_{orb} = 10$ hours, and $e_{t} = 0.3$ is $2.17~{\rm deg ~ yr^{-1}}$, which is of the same order as expected from DNS systems, e.g. smaller than that of PSR J0737-3039 ($4.78~{\rm deg ~ yr^{-1}}$), but larger than PSR B1913+16 ($1.21~{\rm deg ~ yr^{-1}}$). This precession contributes to the observed change in the projected semi-major axis ($x= a_{p} \sin i / c$) and the eccentricity \citep{lk05}. This change in $x$ is an addition to the possible change caused through the change of $i$ caused directly by the spin-orbit coupling (Eqn. 5.15 of DS88).

Several other manifestations of general relativity are effective in case of NSBH binaries, like frame-dragging effect \citep{wk99} and tidal interaction \citep[and references therein]{{dn10}}, etc. Study of these effects was not the aim of this paper.

\section{Conclusions} 
\label{lb:con}

In this work, we have shown that the 2PN and the leading order spin-orbit coupling terms can have significant contributions to the observed value of the periastron advance of an NSBH binary, while the 3PN and higher order terms can be neglected. It will not be difficult to incorporate $k_{\rm 2PN}$ term in the future timing code, as this has a relatively simpler form. On the other hand, $k_{\rm SO}^{\rm LO}$ might be difficult to incorporate as it involves few difficult-to-know terms like the moment of inertia of the neutron star, spin of the black hole and orientations of the vectors. Moreover, higher order terms will be important for modelling gravitational wave emission from such systems. So further theoretical study in this field is necessary before the discovery of the first NSBH binary.

\section*{Acknowledgements}

This work is supported by WVEPSCoR. The author thanks Duncan Lorimer for discussions and the anonymous referee for useful comments.

\end{document}